\documentclass[
    ,final            
  ]
  {aipproc}

\layoutstyle{8x11double}

\newcommand{\eq}[1]{eq.~(\ref{#1})}

\newcommand{\Eq}[1]{Eq.~(\ref{#1})}

\newcommand{\ur}[1]{(\ref{#1})}

\newcommand{\beq}{\begin{equation}}
\newcommand{\eeq}{\end{equation}}

\newcommand{\la}[1]{\label{#1}}
\newcommand{\bea}{\begin{eqnarray}}
\newcommand{\eea}{\end{eqnarray}}
\newcommand{\ba}{\begin{array}}
\newcommand{\ea}{\end{array}}

\newcommand{\half}{{\textstyle{\frac{1}{2}}}}

\newcommand{\n}{\nonumber}
\newcommand{\Tr}{{\rm Tr\,}}

\begin{document}

\title{Statistical physics of dyons and quark confinement}



\author{Dmitri Diakonov}{
  address={Petersburg Nuclear Physics Institute, Gatchina 188300, St. Petersburg, Russia}
}

\author{Victor Petrov}{
  address={Petersburg Nuclear Physics Institute, Gatchina 188300, St. Petersburg, Russia}
}

\begin{abstract}
We present a semiclassical approach to the $SU(N)$ Yang--Mills theory
whose partition function at nonzero temperatures is approximated
by a saddle point -- an ensemble of an infinite number of interacting dyons
of $N$ kinds. The ensemble is governed by an exactly solvable $3d$ quantum
field theory, allowing calculation of correlations functions relevant to confinement.
We show that known criteria of confinement are satisfied in this
semiclassical approximation: (i)~the average Polyakov line is zero below
some critical temperature, and nonzero above it, (ii)~a quark-antiquark pair
has linear rising potential energy, (iii)~the average spatial Wilson loop falls off
exponentially with the area, (iv)~$N^2$ gluons are canceled out from the
spectrum, (v) the critical deconfinement temperature is in good agreement
with lattice data.

Using the same approximation, we find confinement for the exceptional gauge
group $G(2)$ and a first-order deconfinement transition, also in agreement with
lattice findings.
\end{abstract}

\maketitle


\section{Introduction}

Quark confinement is one of the most puzzling phenomena in modern physics.
It is widely believed (and supported by numerical simulations on a lattice)
that if one attempts to separate a probe quark from an antiquark, or
a quark from two other quarks in a proton, a force of about 14 tons
(called the string tension) pulls it back, and this force does not decrease
with the separation! Although long-range forces can be found
in condensed matter physics ({\it e.g.} in the Ising model), it is for
the first time in history we encounter such a phenomenon in a local
microscopic theory, and it cries for an explanation. However, after
35 years~\cite{GMFM} of Quantum Chromodynamics (QCD) not only an {\it ab initio}
mathematical derivation of the ``14 tons'' is absent but there is still no consensus
on what is the qualitative mechanism of quark confinement.

Today we know for sure that nuclear and subnuclear physics is governed by QCD
-- a local renormalizable non-Abelian gauge theory of ``colored'' quarks
interacting with gluons being analogs of photons. Therefore,
the problem of confinement is, at least, formulated mathematically:
it must be a property of the quantum Yang--Mills (YM) theory.
Non-Abelian gauge invariance requires that gluons interact, even in the absence
of quarks. In this paper we consider the pure YM theory with no dynamical quarks,
also called the ``pure glue'' theory. However, quark sources or probes can be
inserted to test certain YM correlation functions of interest, in particular
we shall be interested in the correlation function of Polyakov lines and
the average of large Wilson loops: these are quantities that measure the force
between probe quarks.

Gluons are massless due to gauge invariance, and the only freedom a theorist
has in a pure glue theory is the choice of the gauge group. It is $SU(3)$
for the real world, but it is helpful to consider Yang--Mills theories based
on other Lie groups, in particular $SU(N)$ at any $N$. (At the end of the paper
we consider also the exceptional $G(2)$ group for reasons explained there.)
The scaling of most physical observables with $N$ at large $N$ can be found
from simple $N$-counting rules, therefore considering arbitrary $SU(N)$ gauge
groups provides a powerful check.

The gauge coupling constant $\alpha_s$, the analog of the QED coupling
$\alpha\simeq 1/137$, is in fact not a constant and hence not a parameter
that one can choose at will: it ``runs'' as function of the characteristic
momentum at hand. Owing to asymptotic freedom, it is small at large
and large at small momenta. As in any other asymptotically free theory
with no explicit scale parameter, the ``dimensional transmutation'' occurs
in the YM theory: an exponentially large correlation length $\xi$ appears,
being the renormalization-invariant combination of the ultraviolet cutoff
({\it e.g.} the lattice spacing $a$) and the bare coupling constant $\alpha_s(a)$
given at that cutoff,
\beq
\xi\equiv\frac{1}{\Lambda} = a\,\exp\left(\frac{3}{11}\,\frac{1}{\lambda(a)}\right),
\qquad \lambda(a)=\frac{\alpha_s(a) N}{2\pi}
\la{Lambda}\eeq
where $\lambda(a)$ is the so-called 't Hooft bare coupling constant; it does
not depend on $N$ at large $N$. $\Lambda$ has the dimension of mass; it is
called the YM scale parameter and it actually defines all dimensional
quantities in the theory. The deconfinement phase transition temperature $T_c$
is proportional to $\Lambda$ and the string tension is proportional to $\Lambda^2$
by dimensions. All dimensionless quantities are, generally speaking, of the
order of unity, hence it is a strong-coupling problem from the start.
It makes the theorist's life hard.

Unfortunately, QCD will hardly be ever proved to be an exactly solvable quantum
field theory, even in the large $N$ limit. Therefore, one can either do exact
calculations in a theory that has more symmetries but is not our world
({\it e.g.} considering supersymmetric versions of QCD), or work with QCD
but make approximations. The first is useful as a theoretical laboratory,
the second is necessary to understand semi-quantitatively the key phenomena,
to explain experimental data, and to make predictions.

An approximation is considered to be legitimate if there is a systematic way
of improving its accuracy. The semiclassical approach which we develop below,
belongs to this category. One chooses a saddle-point classical field and then
has to take into account quantum fluctuations about it. Part of the fluctuations
are ultra-violet and are thus the same as in empty space. Their
role is to renormalize the bare coupling constant; at this point the
YM scale parameter \ur{Lambda} emerges. What is left, is a series in
't Hooft's {\it running} coupling $\lambda$ coming from loop expansion
in the background of classical configurations. The running coupling is
evaluated at the maximal momentum in the problem, be it temperature or
the average density of classical configurations. Speaking generally, such
expansion parameter is of the order of unity, however numerically it turns
out to be small: $\lambda$ is between $\frac{1}{4}$ at zero temperature and
$\frac{1}{7}$ near $T_c$. Therefore, in the whole range of temperatures within
the confining phase the semiclassical approximation is expected to yield
an accuracy of 15-25\%, already in the 1-loop approximation (provided the
saddle point is chosen correctly!) with a potential for rapid
improvement when higher loops are taken into account. We shall see, however,
that the actual accuracy can be much better than this estimate. It is not
a too big price to pay if confinement, the most challenging riddle in 35 years,
is explained in simple terms.

We consider the pure Yang--Mills theory based on the $SU(N)$ gauge
group in a broad range of temperatures between 0 and $T_c$, the deconfinement
phase transition temperature. Although the formalism we use is designed for
nonzero $T$, we shall see that the physical observables we find (such as the string
tension) have a finite limit when $T\to 0$.

Confinement, as we understand it today and learn from lattice experiments with
a pure glue theory, has in fact many facets, and all have to be explained.
For example, in a general $SU(N)$ group one can consider ``quarks'' in
various irreducible representations. From the confinement viewpoint all
representations are characterized by the phase it acquires under the
gauge transformation from the group center. The representation is said to
have ``$N$-ality $=k$'' if the phase is $\frac{2\pi k}{N}$.
Let us formulate mathematically the main confinement requirements that
need to be satisfied:
\begin{itemize}
\item the average Polyakov line in any $N$-ality nonzero representation
of the $SU(N)$ group is zero below $T_c$ and nonzero above it
\item the potential energy of two static color sources (defined through
the correlation function of two Polyakov lines) asymptotically rises linearly
with the separation; the slope called the string tension depends only on
the $N$-ality of the sources
\item the average of the spatial Wilson loop decays exponentially with
the area spanning the contour; at vanishing temperatures the spatial (``magnetic'') string
tension has to coincide with the ``electric'' one, for all representations
\item the mass gap: no massless gluons should be left in the spectrum.
\end{itemize}

Remarkably, all these requirements are satisfied already in a semiclassical
approximation if one uses an ensemble of dyons as a saddle point in the Yang--Mills
partition function~\cite{DP-07}.

\section{Yang--Mills theory at nonzero temperatures}

Following Feynman, the Yang--Mills (YM) partition function
can be written as a functional integral over the YM 4-potentials $A_\mu(t,{\bf x})$
that are traceless hermitian $N\!\times\!N$ matrices, satisfying periodic boundary
conditions in imaginary (or Euclidean) time:
\bea\la{Z2}
{\cal Z}\!\!\!\!&\!\!\!\!=\!\!\!\!&\!\!\!\!\!\!\int\!\!DA_\mu(t,{\bf x})\,
\exp\left(\!\!-\frac{1}{2g^2}\!\!\int_0^{\frac{1}{T}}\!\!\!\!dt\!\!\!\int\!\!\!d^3{\bf x}\,
\Tr\,F_{\mu\nu}F_{\mu\nu}\!\right)\!\!,\\
\n
&&A_\mu\left(t+\frac{1}{T},{\bf x}\right)=A_\mu(t,{\bf x}),\;\;T={\rm temperature},
\eea
where $F_{\mu\nu}=\partial_\mu A_\nu-\partial_\nu A_\mu-i[A_\mu A_\nu]$
is the YM field strength.

\begin{figure}[t]
\includegraphics[width=0.40\textwidth]{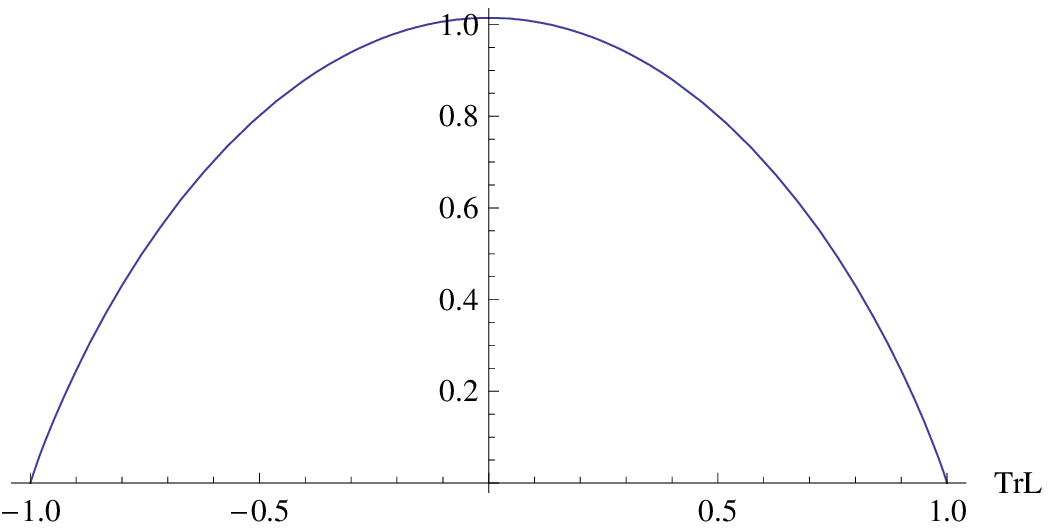} 
\includegraphics[width=0.40\textwidth]{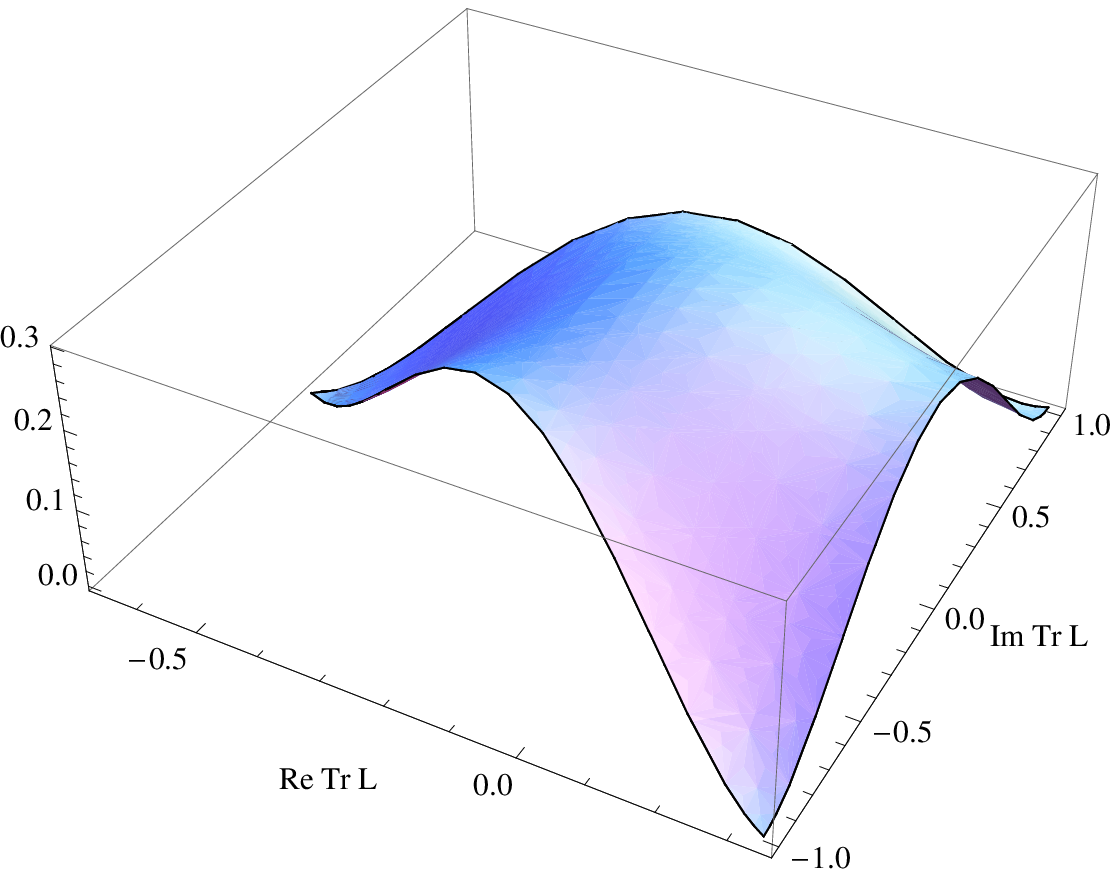} 
\caption{The perturbative potential energy
as function of the Polyakov line for the $SU(2)$ ({\it left}) and $SU(3)$ ({\it right}) groups.
It has minima where the Polyakov loop is one of the $N$ elements
of the center $Z_N$ and is maximal at the ``confining'' holonomy.}
\end{figure}
\begin{figure}[t]
\includegraphics[width=0.40\textwidth]{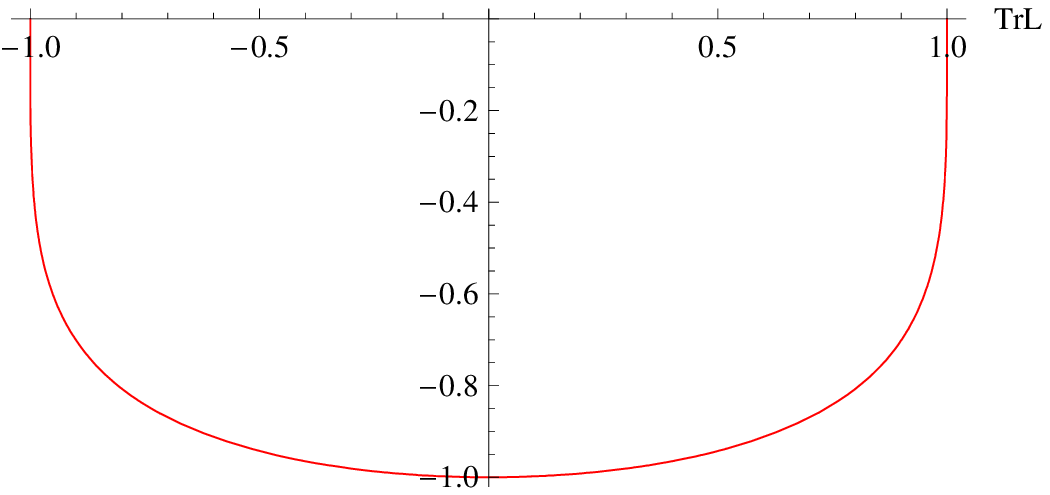}  
\includegraphics[width=0.40\textwidth]{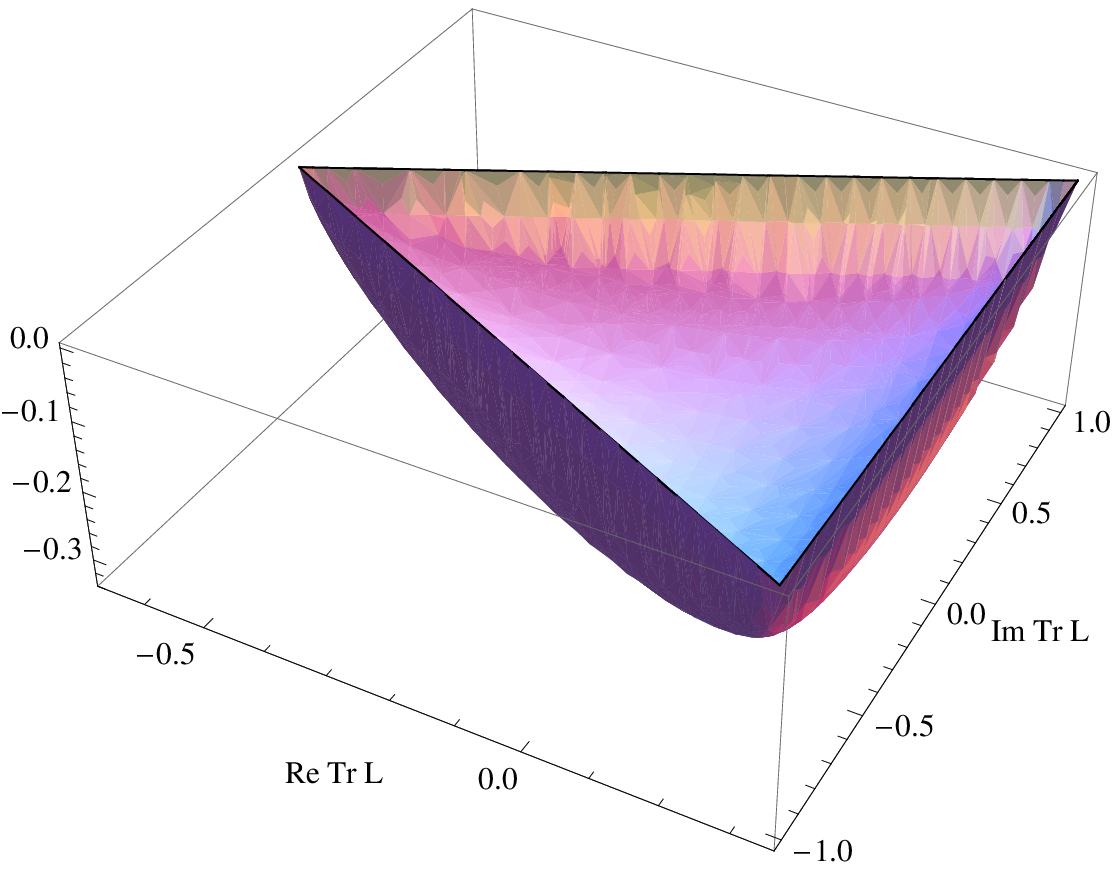}  
\caption{The dyon-induced nonperturbative potential energy
as function of the Polyakov line for the $SU(2)$ ({\it left}) and $SU(3)$ ({\it right}) groups.
Contrary to the perturbative potential energy, it has a single and non-degenerate
minimum at the confining holonomy corresponding to $\Tr L=0$.}
\end{figure}

An important variable is the Polyakov loop; it is the path-ordered exponent in
time direction, hence it can depend only on the space point ${\bf x}$:
\beq
L({\bf x})={\cal P}\,\exp\left(i\int_0^{\frac{1}{T}}\!dt\,A_4(t,{\bf x})\right),
\qquad L\in SU(N).
\la{L}\eeq
Under space-dependent gauge transformations it transforms
as $L\to U^{-1}LU$. The eigenvalues of $L({\bf x})$
are gauge invariant; we parameterize them as
\beq
L={\rm diag}\left(e^{2\pi i \mu_1},e^{2\pi i \mu_2},\ldots ,e^{2\pi i \mu_N}\right),
\la{mu}\eeq
$\mu_1\!+\!\ldots\!+\mu_N=0$, and assume that the phases of these eigenvalues are ordered:
$\mu_1\leq\mu_2\leq\ldots\leq\mu_N\leq\mu_{N+1}\equiv\mu_1\!+\!1$.
We shall call the set of $N$ phases $\{\mu_m\}$ the ``holonomy'' for short.
Apparently, shifting $\mu$'s by integers does not change the eigenvalues, hence all
quantities have to be periodic in all $\mu$'s with a period equal to unity.

The holonomy is said to be ``trivial'' if $L$ belongs to one of the $N$ elements
of the group center $Z_N$. For example, in $SU(3)$ the three trivial holonomies
are
\bea\n
&&\mu_1=\mu_2=\mu_3=0\quad\Longrightarrow\quad
L={\bf 1}_3,\\
\n
&&\mu_1=-\frac{2}{3},\mu_2=\frac{1}{3},\mu_3=\frac{1}{3}\quad\Longrightarrow\quad
L=e^{\frac{2\pi i}{3}}\,{\bf 1}_3,\\
\n
&&\mu_1=-\frac{1}{3},\mu_2=-\frac{1}{3},\mu_3=\frac{2}{3}\quad\Longrightarrow\quad
L=e^{-\frac{2\pi i}{3}}\,{\bf 1}_3.
\eea
Trivial holonomy corresponds to equal $\mu$'s, {\it modulo} unity. Out of all possible
combinations of $\mu$'s a distinguished role is played by {\em equidistant} $\mu$'s
corresponding to $\Tr L=0$:
\beq
\mu_m^{\rm conf}=-\frac{1}{2}-\frac{1}{2N}+\frac{m}{N}.
\la{muconf}\eeq
For example, in $SU(3)$ it is
\beq
\mu_1=-\frac{1}{3},\mu_2=0,\mu_3=\frac{1}{3}\Longrightarrow
L={\rm diag}\left(e^{-\frac{2\pi i}{3}},1,e^{\frac{2\pi i}{3}}\right).
\la{muconf3}\eeq
We shall call it ``most non-trivial'' or ``confining'' holonomy as it
corresponds to $\Tr L=0$ which is the 1$^{\rm st}$ confinement requirement.

Immediately, an interesting question arises: Imagine we take the YM partition
function \ur{Z2} and integrate out all degrees of freedom except
the eigenvalues $\{\mu_m\}$ of the Polyakov loop $L({\bf x})$
which, in addition, we take slowly varying in space. What is the effective action
for $\mu$'s? What set of $\mu$'s is preferred dynamically by the YM system of fields?

First of all, one can address this question in perturbation theory: the result
for the potential energy as function of $\mu$'s is~\cite{GPY,NW}
\beq
P^{\rm pert}\!=\!
\left.\frac{(2\pi)^2T^3}{3}\!\sum_{m>n}^N\!
(\mu_m\!-\!\mu_n)^2[1\!-\!(\mu_m\!-\!\mu_n)]^2\!\right|_{{\rm mod}\,1}\!\!\!\!.
\la{Ppert}\eeq
It is proportional to $T^3$ (by dimensions) and has exactly $N$ zero minima
when all $\mu$'s are equal {\it modulo} unity, see Fig.~1. Hence, $P^{\rm pert}$
says that at least at high temperatures the system prefers one of the
$N$ trivial holonomies corresponding to the Polyakov loop being one of the
$N$ elements of the center $Z_N$. However, terms with gradients of $\mu$'s in
the effective action become negative near ``trivial'' holonomy, signalling
its instability even in perturbation theory~\cite{DO}.

It is interesting that in the supersymmetric ${\cal N}\!=\!1$ version
of the YM theory (where in addition to gluons there are spin-$\half$ gluinos
in the adjoint representation) the perturbative potential energy \ur{Ppert}
is absent in all orders owing to fermion-boson cancelation, but the
nonperturbative potential energy is nonzero. Moreover, it is known exactly
as function of $\mu$'s~\cite{DHKM-99}: it has a single minimum at precisely
the ``most non-trivial'' or ``confining'' holonomy \ur{muconf}.
The result can be traced to the semiclassical contribution of
dyons, which turns out to be exact owing to supersymmetry.

In the non-supersymmetric pure YM theory, the dyon-induced contribution
cannot be computed exactly but only in the semiclassical approximation (this is
what the paper is about), and the perturbative contribution \ur{Ppert} is present, too.
We shall show below that a semiclassical configuration -- an ensemble of dyons with quantum
fluctuations about it -- generates a nonperturbative free energy shown in Fig.~2.
It has the opposite behavior of the perturbative one, having
the minimum at the equidistant (confining) values of the $\mu$'s. There is a fight between
the perturbative and nonperturbative contributions to the free energy~\cite{D-02}.
Since the perturbative contribution to the free energy is $\!\sim\!T^4$ with respect to the nonperturbative one,
it certainly wins when temperatures are high enough, and the system is then forced into one
of the $N$ vacua thus breaking spontaneously the $Z_N$ symmetry. At low temperatures the
nonperturbative contribution prevails forcing the system into the confining vacuum.
At a critical $T_c$ there is a confinement-deconfinement phase transition. It turns out to
be of the second order for $N\!=\!2$ but first order for $N\!=\!3$ and higher, in agreement
with lattice findings.

\section{Dyon saddle points}

Dyons or Bogomolny--Prasad--Sommerfield (BPS) monopoles~\cite{BPS} are (anti) self-dual solutions
of the nonlinear Maxwell equations, $D_\mu^{ab}F^b_{\mu\nu}=0$. In $SU(N)$ there are exactly $N$
kinds of `fundamental' dyons with Coulomb asymptotics for both electric and magnetic fields
(hence the term ``dyon''):
\beq
\pm{\bf E}={\bf B}\stackrel{|{\bf x}|\!\to\!\infty}{=}\frac{1}{2}\frac{{\bf x}}{|{\bf x}|^3}
\times\left\{\begin{array}{c}{\rm diag}(1,-1,0,...,0,0)\\
{\rm diag}(0,1,-1,...,0,0)\\
\ldots\\
{\rm diag}(0,0,0,...,1,-1)\\
{\rm diag}(-1,0,0,...,0,1)\end{array}\right..
\la{EB}\eeq
Dyon solutions are labeled by the holonomy or the set of $\mu$'s at spatial infinity:
\beq
A_4(|{\bf x}|\!\to\!\infty)\to 2\pi T
{\rm diag}(\mu_1,\mu_2,\ldots ,\mu_N).
\la{A4}\eeq
The explicit expressions for the solutions in various gauges can be found {\it e.g.}
in the Appendix of Ref.~\cite{DP-SUSY}.
Inside the cores which are of the size $\sim 1/(T\nu_m)$, the fields are large,
nonlinearity is essential. The action density is time-independent everywhere and is
proportional to the temperature. Isolated dyons are thus $3d$ objects but with finite
action independent of temperature:
\beq
S_{\rm dyon}=\frac{2\pi}{\alpha_s}\nu_m,\quad \nu_m\equiv\mu_{m+1}-\mu_m,\quad \sum_m\nu_m=1,
\la{dyon-action}\eeq
(here $\mu_{N+1}\equiv\mu_1+1$).
The full action of all $N$ kinds of well-separated dyons together is that of one standard
instanton: $S_{\rm inst}=2\pi/\alpha_s$.

In the semiclassical approach, one has first of all to find the statistical weight with which
a given classical configuration enters the partition function. It is given by $\exp(-{\rm Action})$,
times the determinant$^{-1/2}$ from small quantum oscillations about the saddle point. For
an isolated dyon as a saddle-point configuration, this factor diverges linearly in the infrared
region owing to the slow Coulomb decrease of the dyon field \ur{EB}. It means that isolated
dyons are not acceptable as saddle points: they have zero weight, despite finite classical
action. However, one may look for classical solutions that are superpositions of $N$ fundamental
dyons, with zero net magnetic charge. The small-oscillation determinant must be infrared-finite
for such classical solutions, if they exist.

\section{Instantons with non-trivial holonomy}
\begin{figure}[t]
\includegraphics[width=0.32\textwidth]{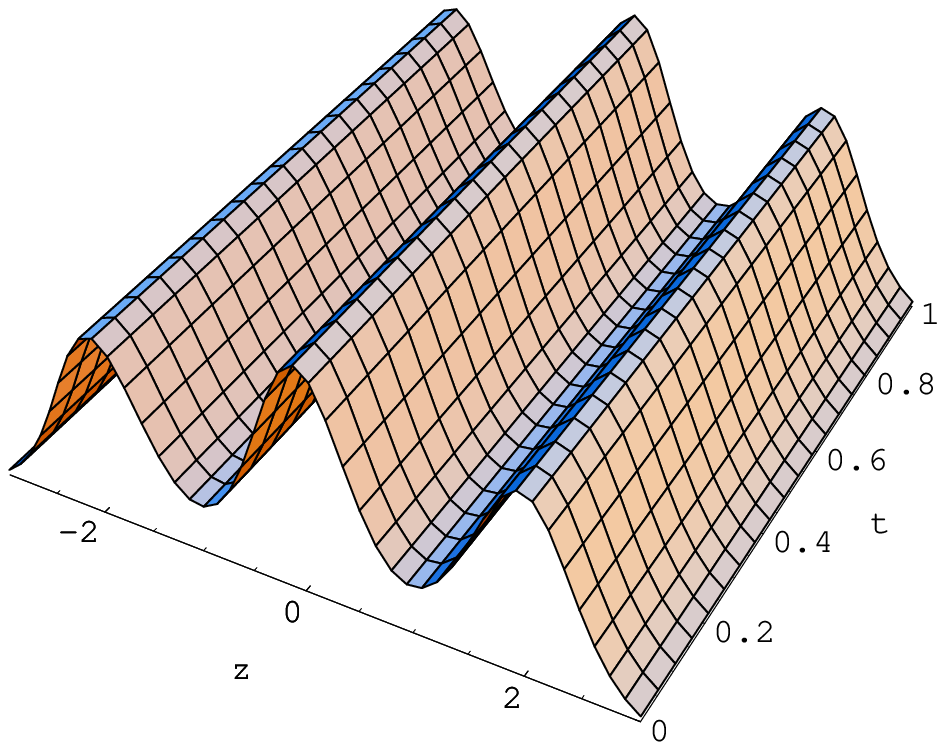} 
\includegraphics[width=0.32\textwidth]{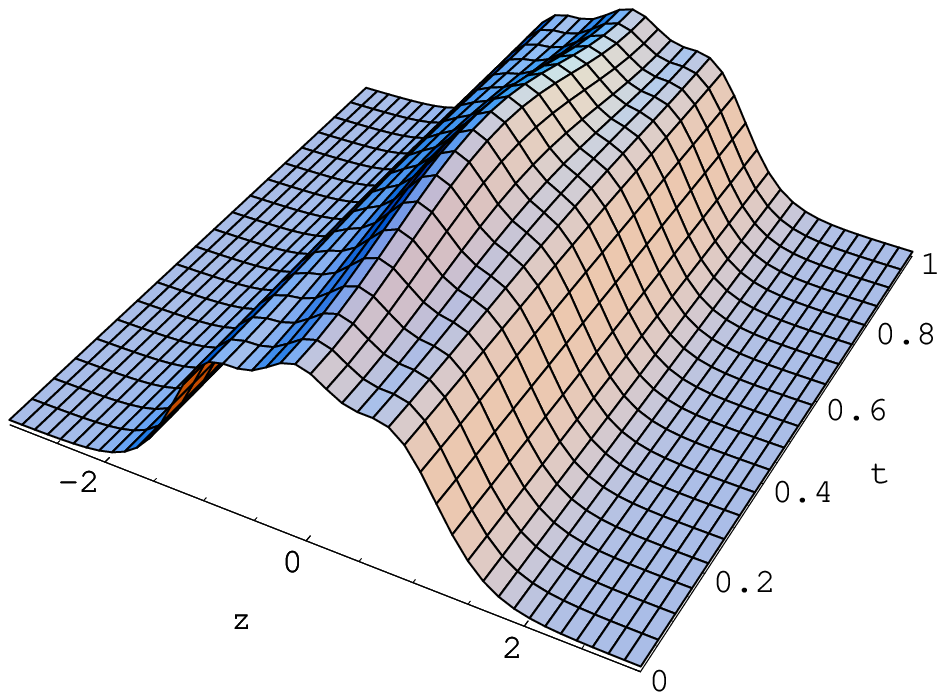} 
\includegraphics[width=0.32\textwidth]{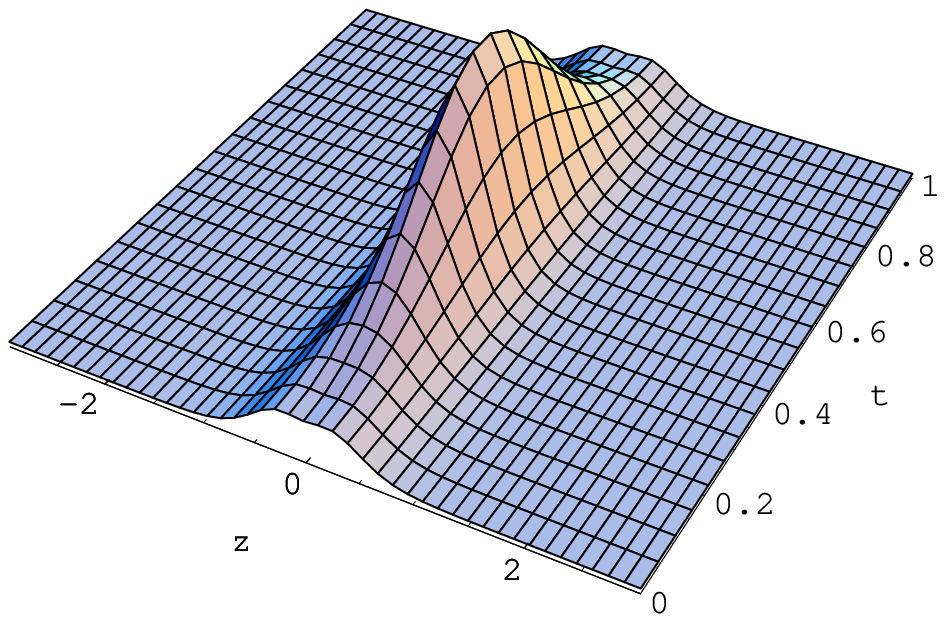}  
\caption{Action density inside the $SU(3)$ KvBLL instanton as function of time and
one space coordinate, for large ({\it left}), intermediate ({\it middle}) and small
({\it right}) separations between the three constituent dyons. }
\end{figure}

The needed classical solution has been found a decade ago by Kraan
and van Baal~\cite{KvB} and independently by Lee and Lu~\cite{LL},
see also~\cite{LeeYi}. We shall call them for short the ``KvBLL instantons''; an
alternative name is ``calorons with nontrivial holonomy''. The solution was first found
for the $SU(2)$ group but soon generalized to an arbitrary $SU(N)$~\cite{KvBSUN},
see~\cite{vBZak} for a review.

The general solution $A_\mu^{\rm KvBLL}$ depends on Euclidean time $t$ and
space ${\bf x}$ and is parameterized by $3N$ positions of $N$ kinds of
`constituent' dyons in space ${\bf x}_1,\ldots, {\bf x}_N$
and their $U(1)$ phases $\psi_1,\ldots,\psi_N$. All in all, there are $4N$ collective
coordinates characterizing the solution (called the moduli space), of which the action
$S_{\rm inst}=2\pi/\alpha_s$ is in fact independent, as it should be for a general
solution with a unity topological charge. The solution also depends explicitly on
temperature $T$ and on the holonomy $\mu_1,\ldots,\mu_N$:
\beq
A_\mu^{\rm KvBLL}=\bar A_\mu^a(t, {\bf x};\, {\bf x}_1,\ldots, {\bf x}_N,\psi_1,\ldots,\psi_N;\,
T,\mu_1,\ldots,\mu_N).
\la{KvBLLi}\eeq
The solution is a relatively simple expression given by elementary functions.
If the holonomy is trivial (all $\mu$'s are equal {\it modulo} unity) the expression
takes the form of the strictly periodic $O(3)$ symmetric caloron~\cite{HS} reducing further
to the standard $O(4)$ symmetric BPST instanton~\cite{BPST} in the $T\to 0$ limit.
At small temperatures but arbitrary holonomy, the KvBLL instanton also has only a small
${\cal O}(T)$ difference with the standard instanton.

One can plot the action density of the KvBLL instanton in various corners of the
parameter (moduli) space, see Fig.~3.

When all dyons are far apart one observes $N$ static ({\it i.e.} time-independent)
objects, the isolated dyons. As they merge, the configuration is not static anymore,
it becomes a {\it process} in time. In the limiting case of a complete merger, the
configuration becomes a $4d$ lump resembling the standard instanton. The full
(integrated) action is exactly the same $S_{\rm inst}=2\pi/\alpha_s$ for any choice
of the dyon separations. It means that classically dyons do not interact. However,
they do experience a peculiar interaction at the quantum level to which we proceed.

\section{Quantum weight of many dyons}

Remarkably, the small-oscillation determinant about a single KvBLL $SU(N)$
instanton made of $N$ different-kind dyons can be computed exactly~\cite{DGPS,S}.
With this experience, the quantum weight of an arbitrary number of dyons of
$N$ kinds has been suggested in Ref.~\cite{DP-07}. In the YM partition function,
there are saddle points corresponding to any set of $K_m$ dyons.
In the thermodynamic limit $V\to\infty$ one needs to take a saddle point with
${\cal O}(V)$ dyons. Let $K_m$ be the number of dyons of kind $m$ ($m=1\ldots N$)
and let ${\bf x}_{mi}$ be the coordinate of the $i^{\rm th}$ dyon of kind $m$
($i=1\ldots K_m$). In the semiclassical approximation the YM partition function
is approximated by the partition function of a grand canonical ensemble of
$K_1+K_2+\ldots +K_N$ dyons,
\beq
{\cal Z}=\sum_{K_1...K_N}\frac{1}{K_1!...K_N!}\prod_{m=1}^N\,\prod_{i=1}^{K_m}
\int (d{\bf x}_{mi}\,f)\,\sqrt{\det g({\bf x}_{mi})},
\la{Z3}\eeq
where $g({\bf x}_{mi})$ is a $4(K_1+\ldots +K_N)\times 4(K_1+\ldots +K_N)$
metric tensor of the dyons' moduli space, composed by the overlaps of zero modes
of individual dyons, and $f$ is the fugacity,
\beq
f=\frac{N^2}{16\pi^3\lambda^2}\,\frac{\Lambda^4}{T}={\cal O}(N^2).
\la{f}\eeq
The bare 't Hooft coupling constant $\lambda$ is renormalized and starts to ``run''
only at the 2-loop level not considered here. Eventually, its argument will be
the largest scale in the problem, be it the temperature or the equilibrium density
of dyons.

It is not difficult to find the metric tensor $g({\bf x}_{mi})$ for well-separated dyons.
In this case the four zero modes $\phi^{(\kappa)}_\mu (\kappa=1,2,3,4)$ of individual dyons
are given by the components of the field strength: $\phi^{(\kappa)}_\mu=F_{\mu\kappa}$.
The zero modes for the $m^{\rm th}$ kind of dyon are normalized to its action,
$\int \Tr \phi^{(\kappa)}_\mu\phi^{(\lambda)}_\mu\sim\delta^{\kappa\lambda}\nu_m$
(see \Eq{dyon-action}) and hence depend on the holonomy. Since the field strengths
decay as $1/r^2$ (see \Eq{EB}) the overlaps between zero modes are Coulomb-like,
and only those that are nearest neighbors in $m$ do interact. In fact, the diagonal
components of the metric tensor also acquire Coulomb-like corrections since the action
of individual dyons is actually normalized to its asymptotic field $A_4$ that gets
Coulomb corrections from other dyons.

As a result, we obtain the $4(K_1+\ldots +K_N)\times 4(K_1+\ldots +K_N)$ metric tensor
$g({\bf x}_{mi})$ with Coulomb interactions as entries, and the $\nu_m$'s on the diagonal.
It turns out that its determinant is a square of the determinant of a related matrix,
$\sqrt{\det g}=\det G$ where $G$ is a $(K_1+\ldots +K_N)\times (K_1+\ldots +K_N)$ matrix:
\beq\la{G}
G_{mi,nj}=\delta_{mn}\delta_{ij}\,\left(4\pi\nu_m
+\sum_k\frac{1}{T|{\bf x}_{mi}\!-\!{\bf x}_{m-1,k}|}\right.
\eeq
$$
+\left.\sum_k\frac{1}{T|{\bf x}_{mi}\!-\!{\bf x}_{m+1,k}|}
-2\sum_{k\neq i}\frac{1}{T|{\bf x}_{mi}\!-\!{\bf x}_{mk}|}\right)
$$
$$
-\frac{\delta_{m,n-1}}{T|{\bf x}_{mi}\!-\!{\bf x}_{m+1,j}|}
-\frac{\delta_{m,n+1}}{T|{\bf x}_{mi}\!-\!{\bf x}_{m-1,j}|}
+2\left.\frac{\delta_{mn}}{T|{\bf x}_{mi}\!-\!{\bf x}_{mj}|}\right|_{i\neq j},
$$
where ${\bf x}_{mi}$ is the coordinate of the $i^{\rm th}$ dyon of kind $m$.
The matrix $G$ has the following nice properties:
\begin{itemize}
\item symmetry: $G_{mi,nj}=G_{nj,mi}$
\item overall ``neutrality'': the sum of Coulomb interactions in non-diagonal entries
cancel those on the diagonal:
$\sum_{nj}G_{mi,nj}=4\pi\nu_m$
\item identity loss: dyons of the same kind are indistinguishable, meaning
mathematically that $\det G$ is symmetric under permutation of any pair of dyons
$(i\!\leftrightarrow\!j)$ of the same kind $m$. Dyons do not `know' to which
instanton they belong to
\item attraction/repulsion: if one decreases the separation between same-kind dyons
or increases the separation between different-kind dyons, the $\det G$ decreases.
It means that same-kind dyons repulse each other whereas different-kind dyons
attract each other. The $\det G$ measure favors formation of neutral clusters
with $N$ different kinds of dyons
\item factorization: in the geometry when dyons fall into $K$ well separated
neutral clusters of $N$ dyons of different kinds, $\det G$ factorizes
into a product of {\it exact} integration measures for $K$ KvBLL instantons~\cite{Kraan,DG-05}
valid for {\it any} separations between different-kind dyons,
including their strong overlap
\item last but not least, the metric $g$ corresponding to $G$ is hyper-K\"ahler,
as it should be for the moduli space of a self-dual classical field~\cite{AH}.
In fact, it is a severe restriction on the metric.
\end{itemize}
An overall constant factor depending on the holonomy and temperature,
$\exp\left(-P^{\rm pert}V\right)$, is understood in \Eq{Z3}, where $P^{\rm pert}$
is the perturbative gluon loop \ur{Ppert} in the background of a
constant field $A_4$ \ur{A4}. This factor arises from the non-zero modes
in the fluctuation determinant about dyons and is necessarily present
as most of the $3d$ space outside the dyons' core is just a constant $A_4$ background.
Indeed the calculation~\cite{DGPS,S} exhibits this factor which is the only one
proportional to the 3-volume $V$.

The ensemble defined by a determinant of a matrix whose dimension is the number of
particles, is not a usual one. More customary, the interaction is given by
the Boltzmann factor $\exp\left(-U_{\rm int}({\bf x}_1,\ldots)\right)$. Of course,
one can always present the determinant in that way using the identity
$\det G=\exp(\Tr\log G)\equiv \exp(-U_{\rm int})$ but the interactions will then
include three-, four-, five-... body forces. At the same time, it is precisely
the determinant form of the interaction that makes the statistical physics of dyons
an exactly solvable problem.

\section{Statistical physics of dyons as a Quantum Field Theory}

It is possible to present the grand canonical ensemble of dyons, governed by the
interaction \ur{G} as an equivalent $3d$ quantum field theory. This will
enable us to compute various correlation functions of interest.

To proceed to the quantum field theory description we use two mathematical tricks.\\

1. \underline{``Fermionization''} (Berezin~\cite{Berezin}). It is helpful to exponentiate
the Coulomb interactions rather than keeping them in $\det G$.
To that end one presents the determinant of a matrix as an integral over a finite
number of anticommuting Grassmann variables ($\{\psi_A^\dagger\,\psi_B\}=\delta_{AB}$):
$$
\det (G_{AB})
=\int\!\prod_A d\psi_A^\dagger\,d\psi_A\,
\exp\left(\psi_A^\dagger\,G_{AB}\,\psi_B\right)\,.
$$

Now we have the two-body Coulomb interactions in the exponent and it is possible
to use the second trick.\\

2. \underline{``Bosonization''} (Polyakov~\cite{Polyakov77}). One can present
the Coulomb interactions in the exponent with the help of a Gaussian integral
over an auxiliary field $\phi$:
\bea\n
&&\exp\left(\sum_{m,n}\frac{Q_mQ_n}{|{\bf x}_m-{\bf x}_n|}\right)
=\int D\phi\,\exp\left[-\int\!d{\bf x}\right.\\
\n
&&\cdot\left.\left(\frac{1}{16\pi}\partial_i\phi\partial_i\phi
+\rho\phi\right)\right]=\exp\left(\int \rho\frac{4\pi}{\triangle}\rho\right),\\
\n
&&\rho = \sum Q_m\,\delta({\bf x}-{\bf x}_m).
\eea

After applying the first trick the ``charges'' $Q_m$ become Grassmann variables but
after applying the second one, it becomes easy to integrate them out since the square of
a Grassmann variable is zero. In fact one needs $2N$ boson fields ${\rm v}_m,{\rm w}_m$ to reproduce
diagonal elements of $G$ and $2N$ anticommuting (``ghost'') fields $\chi^\dagger_m,\chi_m$
to present the non-diagonal elements. The chain of identities is accomplished in
Ref.~\cite{DP-07} and the result for the partition function for the dyon ensemble \ur{Z3} 
is, identically, a path integral defining a quantum field theory in 3 dimensions:
\bea\n
&&{\cal Z}=\int\!D\chi^\dagger\,D\chi\,D{\rm v}\,D{\rm w}\,\exp\int\!d^3x
\left\{\frac{T}{4\pi}\,\left(\partial_i\chi_m^\dagger\partial_i\chi_m\right.\right.\\
\n
&&+\left.\partial_i{\rm v}_m\partial_i{\rm w}_m\right)+f\left[(-4\pi\mu_m+{\rm v}_m)\frac{\partial{\cal F}}{\partial {\rm w}_m}\right.\\
&&+\left.\left.\chi^\dagger_m\,\frac{\partial^2{\cal F}}{\partial {\rm w}_m\partial {\rm w}_n}\,\chi_n\right]
\right\},\quad
{\cal F}=\sum_{m=1}^N e^{{\rm w}_m-{\rm w}_{m+1}}\,.
\la{Z4}\eea
The fields ${\rm v}_m$ have the meaning of the asymptotic Abelian electric potentials of dyons,
\bea\la{Av}
\left(A_4\right)_{mn}&=&\delta_{mn}\,A_{m\,4},\\
\n
A_{m\,4}({\bf x})/T &=& 2\pi \mu_m-\half {\rm v}_m({\bf x}),\;\;\; {\bf E}_m={\bf \nabla}A_{m\,4},
\eea
while ${\rm w}_m$ have the meaning of the dual (or magnetic) Abelian potentials.
Note that the kinetic energy for the ${\rm v}_m,{\rm w}_m$ fields has only the mixing term
$\partial_i{\rm v}_m\partial_i{\rm w}_m$ which is nothing but the Abelian duality transformation
${\bf E}\cdot{\bf B}$. The function ${\cal F}({\rm w})$ in \ur{Z4} where one assumes
a cyclic summation over $m$, is known as the periodic (or affine) Toda lattice.

Although the Lagrangian in \Eq{Z4} describes a highly nonlinear interacting quantum field
theory, it is in fact exactly solvable! To prove it, one observes that the fields ${\rm v}_m$
enter the Lagrangian only linearly, therefore one can integrate them out. It leads
to a functional $\delta$-function:
\beq
\int\!D{\rm v}_m\quad\longrightarrow\quad \delta\left(-\frac{T}{4\pi}\partial^2{\rm w}_m
+f\frac{\partial{\cal F}}{\partial {\rm w}_m}\right).
\la{UD}\eeq
This $\delta$-function restricts possible fields ${\rm w}_m$ over which one still has
to integrate in \eq{Z4}. Let $\bar {\rm w}_m$ be a solution to the argument of the
$\delta$-function. Integrating over small fluctuations about $\bar{\rm w}$ gives
the Jacobian
\beq\la{Jac}
{\rm Jac}={\rm det}^{-1}\left(-\frac{T}{4\pi}\partial^2\delta_{mn}
+\left.f\frac{\partial^2{\cal F}}{\partial {\rm w}_m\partial {\rm w}_n}\,
\right|_{{\rm w}=\bar{\rm w}}\,\right)\,.
\eeq
Remarkably, exactly the same functional determinant but in the numerator
arises from integrating over the ghost fields, for any background $\bar{\rm w}$.
Therefore, all quantum corrections cancel {\em exactly} between the boson and
ghost fields (a characteristic feature of supersymmetry), and the ensemble of dyons
is basically governed by a classical field theory.

To find the ground state we examine the fields' potential energy being
$-4\pi f\mu_m\partial{\cal F}/\partial {\rm w}_m$ which we prefer to write restoring
$\nu_m=\mu_{m\!+\!1}-\mu_m$ and ${\cal F}$ as
\beq
{\cal P}=-4\pi f V\sum_m \nu_m\,e^{{\rm w}_m-{\rm w}_{m\!+\!1}}
\la{calP1}\eeq
(the volume factor arises for constant fields ${\rm w}_m$). One has first to find the
stationary point in ${\rm w}_m$ for a given set of $\nu_m$'s. It leads to the equations
$$
\frac{\partial {\cal P}}{\partial {\rm w}_m}=0
$$
whose solution is
\beq
e^{w_1-w_2}=\frac{(\nu_1\nu_2\nu_3...\nu_N)^{\frac{1}{N}}}{\nu_1},\quad
{\rm etc.}
\la{extr2}\eeq
Putting it back into \eq{calP1} we obtain
\beq
{\cal P}=-4\pi f V N (\nu_1\nu_2...\nu_N)^{\frac{1}{N}},\quad
\nu_1+...+\nu_N=1.
\la{calP2}\eeq
The minimum equal ${\cal P}_{\rm min}=-4\pi f V$ is achieved at 
$\nu_1=\ldots =\nu_N =\frac{1}{N}$, that is at equidistant, confining 
value of the holonomy, cf. \Eq{muconf}.
We have also proven that the result is exact, as all potential quantum corrections
cancel in the partition function \ur{Z4}.

Given this cancelation, the key finding -- that the dyon-induced free energy has the
minimum at the confining value of holonomy -- is trivial. If all Coulomb interactions cancel
after integration over dyons' positions, the weight of a many-dyon configuration
is the same as if they were infinitely dilute (although they are not).
Then the weight, what concerns the holonomy, is proportional to the product of diagonal
matrix elements of $G$ in the dilute limit, that is to the normalization integrals
for dyon zero modes, that is to the product of the dyon actions $\sim \nu_m$
where $\nu_m=\mu_{m+1}-\mu_m$ and $\nu_N=\mu_1+1-\mu_N$
such that $\nu_1+\nu_2+\ldots+\nu_N=1$. The sum of all $N$ kinds of dyons' actions is
fixed and equal to the instanton action, however, it is the {\em product} of actions
that defines the weight. The product is maximal when all actions are equal, hence
the equidistant or confining $\mu$'s are statistically preferred. Thus, the average
Polyakov line is zero, $<\Tr L>=0$.

\section{Heavy quark potential}

The field-theoretic representation of the dyon ensemble enables one to compute
various YM correlation functions in the semiclassical approximation.
The key observables relevant to confinement are the correlation function
of two Polyakov lines (defining the heavy quark potential), and the average of
large Wilson loops. A detailed calculation of these quantities is performed in Ref.~\cite{DP-07};
here we only present the results and discuss the meaning.

\subsection{$N$-ality and $k$-strings}

From the viewpoint of confinement, all irreducible representations of the $SU(N)$
group fall into $N$ classes: those that appear in the direct product of any number
of adjoint representations, and those that appear in the direct product of any
number of adjoint representations with the irreducible representation being the
rank-$k$ antisymmetric tensor, $k=1,\ldots , N\!-\!1$. ``$N$-ality'' is said
to be zero in the first case and equal to $k$ in the second. $N$-ality-zero
representations transform trivially under the center of the group $Z_N$;
the rest acquire a phase $2\pi k/N$.

One expects that there is no asymptotic linear potential between static color
sources in the adjoint representation as such sources are screened by gluons.
If a representation is found in a direct product of some number of adjoint
representations and a rank-$k$ antisymmetric representation, the adjoint ones
``cancel out'' as they can be all screened by an appropriate number of gluons.
Therefore, from the confinement viewpoint all $N$-ality $=k$ representations are
equivalent and there are only $N-1$ string tensions $\sigma_{k,N}$ being the
coefficients in the {\em asymptotic} linear potential for sources in the antisymmetric
rank-$k$ representation. They are called ``$k$-strings''.

The value $k\!=\!1$ corresponds to the fundamental representation whereas
$k=N\!-\!1$ corresponds to the representation conjugate to the fundamental
[quarks and anti-quarks]. In general, the rank-$(N\!-\!k)$ antisymmetric
representation is conjugate to the rank-$k$ one; it has the same dimension
and the same string tension, $\sigma_{k,N}=\sigma_{N\!-\!k,N}$.

The behavior of $\sigma_{k,N}$ as function of $k$ and $N$ is an important issue as
it discriminates between various confinement mechanisms. On general $N$-counting
grounds one can only infer that at large $N$ and $k\ll N$, $\sigma_{k,N}/\sigma_{1,N}=
(k/N)(1+{\cal O}(1/N^2))$. Important, there should be no ${\cal O}(N^{-1})$
correction~\cite{Shifman}. A popular version called ``Casimir scaling'', according
to which the string tension is proportional to the Casimir operator for a
given representation (it stems from an idea that confinement is somehow related
to the modification of a one-gluon exchange at large distances), does not satisfy
this restriction.

\subsection{Correlation function of Polyakov lines}

To find the potential energy $V_{k,N}$ of static ``quark'' and ``antiquark''
transforming according to the antisymmetric rank-$k$ representation, one has
to consider the correlation of Polyakov lines in the appropriate representation:
\beq
\left<\!\Tr L_{k,N}({\bf z}_1)\;\Tr L^\dagger_{k,N}({\bf z}_2)\!\right>\!
=\!{\rm const.}\,\exp\left(\!-\!\frac{V_{k,N}({\bf z}_1\!-\!{\bf z}_2)}{T}\!\right).
\la{VkN}\eeq
Far away from dyons' cores the field is Abelian and in the field-theoretic language
of \Eq{Z4} is given by \Eq{Av}. Therefore, the Polyakov line in the fundamental
representation is
\beq
\Tr L({\bf z})\!=\!\!\sum_{m=1}^N\! Z_m,\;\; Z_m=\exp\left(2\pi i\mu_m\!-\!\frac{i}{2}{\rm v}_m({\bf z})\right).
\la{Lv}\eeq
In the general antisymmetric rank-$k$ representation
\beq
\Tr L_{k,N}({\bf z})=\sum_{m_1<m_2<...<m_k}^NZ_{m_1}Z_{m_2}...Z_{m_k}
\la{LkNv}\eeq
where cyclic summation from 1 to $N$ is assumed.

The average \ur{VkN} can be computed from the quantum field theory \ur{Z4}.
Inserting the two Polyakov lines \ur{LkNv} into \Eq{Z4} we observe that
the Abelian electric potential ${\rm v}_m$ enters linearly in the exponent as before.
Therefore, it can be integrated out, leading to a $\delta$-function for the
dual field ${\rm w}_m$, which is now shifted by the source (cf. \Eq{UD}):
\bea\n
\int\!D{\rm v}_m &\longrightarrow & \prod_m\delta\left(-\frac{T}{4\pi}\partial^2{\rm w}_m
+f\frac{\partial{\cal F}}{\partial {\rm w}_m}\right.\\
\n
&-&\frac{i}{2}\,\delta({\bf x}\!-\!{\bf z_1})(\delta_{mm_1}+\ldots +\delta_{mm_k})\\
\n
&+&\left.\frac{i}{2}\,\delta({\bf x}\!-\!{\bf z_2})(\delta_{mn_1}+\ldots +\delta_{mn_k})\right).
\la{UDL}\eea
One has to find the dual field ${\rm w}_m({\bf x})$ nullifying the argument of this
$\delta$-function, plug it into the action
\beq
\exp\left(\!\int\!d{\bf x}\,\frac{4\pi f}{N}{\cal F}(w)\right),
\la{action}\eeq
and sum over all sets $\{m_1\!<\!m_2\!<\!...\!<\!m_k\}$, $\{n_1\!<\!n_2\!<\!...\!<\!n_k\}$
with the weight $\exp\left(2\pi i(m_1\!+\!\ldots \!+\!m_k-n_1\!-\!\ldots \!-\!n_k)/N\right)$.
The Jacobian from resolving the $\delta$-function again cancels exactly with the
determinant arising from ghosts. Therefore, the calculation of the correlator \ur{VkN},
sketched above, is exact.

At large separations between the sources $|{\bf z}_1\!-\!{\bf z}_2|$, the fields ${\rm w}_m$
resolving the $\delta$-function are small and one can expand the Toda chain:
\beq
{\cal F}(w)=\sum_m e^{{\rm w}_m-{\rm w}_{m\!+\!1}}\approx N+\frac{1}{2}\, {\rm w}_m\,{\cal M}_{mn}\,{\rm w}_n,
\la{calFsm}\eeq
where
\beq
{\cal M}=\left(\begin{array}{cccccc}2&-1&0&\ldots&0 & -1\\ -1&2&-1&\ldots &0& 0\\
0&-1&2&-1&\ldots &0\\ \ldots & \ldots & \ldots & \ldots & \ldots & \ldots\\-1&0&0&\ldots
&-1&2\end{array}\right).
\la{calM}\eeq
As apparent from \Eq{calFsm}, the eigenvalues of ${\cal M}$ determine the spectrum of the
dual fields ${\rm w}_m$. There is one zero eigenvalue which decouples from everywhere,
and $N\!-\!1$ nonzero eigenvalues
\beq
{\cal M}^{(k)}=\left(2\sin\frac{\pi k}{N}\right)^2,\quad k=1,...,N-1.
\la{eig2}\eeq
Certain orthogonality relation imposes the selection rule: the asymptotics of the correlation
function of two Polyakov lines in the antisymmetric rank-$k$ representation is determined
by precisely the  $k^{\rm th}$ eigenvalue. We obtain~\cite{DP-07}
\bea
&&\left<\Tr L_{k,N}({\bf z}_1)\;\;\Tr L^\dagger_{k,N}({\bf z}_2)\right>\\
\n
&&\stackrel{z_{12}\to\infty}{=}{\rm const.}\,
\exp\left(-|{\bf z}_1-{\bf z}_2|\,M\sqrt{{\cal M}^{(k)}}\!\right)
\la{corrLk2}\eea
where $M$ is the `dual photon' mass,
\beq
M=\sqrt{\frac{4\pi f}{T}}=\frac{N\Lambda^2}{2\pi\lambda T}={\cal O}(N).
\la{M}\eeq
Comparing it with the definition of the heavy quark potential \ur{VkN} we find
that there is an asymptotically linear potential between static ``quarks'' in
any $N$-ality nonzero representation, with the $k$-string tension
\beq
\sigma_{k,N}=MT\sqrt{{\cal M}^{(k)}}=2MT\,\sin\frac{\pi k}{N}
=\frac{\Lambda^2}{\lambda}\,\frac{N}{\pi}\,\sin\frac{\pi k}{N}.
\la{sigma-k}\eeq
This is the so-called `sine regime': it has been found before in certain supersymmetric
theories~\cite{sine}. Lattice simulations~\cite{DelDebbio-k} support this regime,
whereas another lattice study~\cite{Teper-k} gives somewhat smaller values
but within two standard deviations from the values following from \eq{sigma-k}.

We see that at large $N$ and $k\ll N$, $\sigma_{k,N}/\sigma_{1,N}=
(k/N)(1+{\cal O}(1/N^2))$, as it should be on general grounds, and that
all $k$-string tensions have a finite limit at zero temperature.

\section{Area law for large Wilson loops}

The magnetic field of dyons beyond their cores is Abelian
and is a superposition of the Abelian fields of individual dyons. For large
Wilson loops we are interested in, it is this superposition field of a large
number of dyons that contributes most as they have a slowly decreasing
$1/|{\bf x}\!-\!{\bf x}_i|$ asymptotics, hence the use of the field outside
the cores is justified. Owing to self-duality,
\beq
\left[B_i({\bf x})\right]_{mn}=\left[\partial_iA_4({\bf x})\right]_{mn}
=-\frac{T}{2}\,\delta_{mn}\,\partial_i{\rm v}_m({\bf x}),
\la{B}\eeq
cf. \eq{Av}. Since $A_i$ is Abelian beyond the cores, one can use the Stokes
theorem for the spatial Wilson loop:
\bea\n
W&\equiv &\Tr\,{\cal P}\exp\,i\oint\!A_idx^i=\Tr\exp\,i\int\!B_i\,d^2\sigma^i\\
\la{Wi1}
&=&\sum_m\exp\left(-i\frac{T}{2}\int\!d^2\sigma^i\,\partial_i{\rm v}_m\right).
\eea
\Eq{Wi1} may look contradictory as we first use $B_i={\rm curl}\,A_i$ and then
$B_i=\partial_iA_4$. Actually there is no contradiction as the last equation
is true up to Dirac string singularities which carry away the magnetic flux.
If the Dirac string pierces the surface spanning the loop it gives a quantized
contribution $\exp(2\pi i\!\cdot\!{\rm integer})=1$; one can also use the gauge
freedom to direct Dirac strings parallel to the loop surface in which case
there is no contribution from the Dirac strings at all.

Let us take a flat Wilson loop lying in the $(xy)$ plane at $z\!=\!0$. Then
\eq{Wi1} is continued as
\bea\n
W&=&\sum_m\exp\left(-i\frac{T}{2}\int_{x,y\in {\rm Area}}\!d^3x\,\partial_z{\rm v}_m\delta(z)\right)\\
&=&\sum_m\exp\left(i\frac{T}{2}\int_{x,y\in {\rm Area}}\!d^3x\,{\rm v}_m\,\partial_z\delta(z)\right)\,.
\la{Wi2}\eea
It means that the average of the Wilson loop in the dyon ensemble is given by
the partition function \ur{Z4} with the source
$$
\sum_m\exp\left(i\frac{T}{2}\int\!d^3x\;{\rm v}_m\,\frac{d\delta(z)}{dz}\,
\theta(x,y\in {\rm Area})\right)
$$
where $\theta(x,y\in {\rm Area})$ is a step function equal to unity if $x,y$ belong
to the area inside the loop and zero otherwise.

As in the case of the Polyakov lines the presence of the Wilson loop shifts
the argument of the $\delta$-function arising from the integration over
the ${\rm v}_m$ variables, and the ghost determinant cancels exactly the Jacobian
from the fluctuations of ${\rm w}_m$'s, therefore the classical-field calculation
is exact.

One has to solve the non-linear Toda equations on ${\rm w}_m$'s with
a source along the surface of the loop,
\bea\n
&&-\partial^2{\rm w}_m+M^2\left(e^{{\rm w}_m-{\rm w}_{m+1}}-e^{w_{m-1}-{\rm w}_m}\right)\\
&&=-2\pi i\,\delta_{mm_1}\,\frac{d\delta(z)}{dz}\,\theta(x,y\in {\rm Area}),
\la{UD3}\eea
for all $m_1$, plug it into the action $(4\pi f/N){\cal F}(w)$, and sum over
$m_ 1$. In order to evaluate the average of the Wilson loop in a general antisymmetric
rank-$k$ representation, one has to take the source in \eq{UD3} as $-2\pi i\,\delta'(z)\,
\left(\delta_{mm_1}\!+\!\ldots\!+\!\delta_{mm_k}\right)$ and sum over $m_1\!<\!\ldots\!<\!m_k$
from 1 to $N$, see \eq{LkNv}.

Contrary to the case of the Polyakov lines, one cannot, generally speaking,
linearize \eq{UD3} in ${\rm w}_m$ but has to solve the non-linear equations as they are.
The Toda equations \ur{UD3} with a $\delta'(z)$ source in the r.h.s. define
``pinned soliton'' solutions ${\rm w}_m(z)$ that are $1d$ functions in the direction
transverse to the surface spanning the Wilson loop but do not depend on the
coordinates $x,y$ provided they are taken inside the loop. Beyond that surface ${\rm w}_m=0$.
Along the perimeter of the loop, ${\rm w}_m$ interpolate between the soliton and zero.
For large areas, the action \ur{action} is therefore proportional to the area of the surface
spanning the loop, which gives the famous area law for the average Wilson loop.
The coefficient in the area law, the `magnetic' string tension, is found
from integrating the action density of the soliton ${\rm w}_m(z)$ in the $z$ direction.

The exact solutions of \Eq{UD3} for any $N$ and any representation $k$ have been
found in Ref.~\cite{DP-07}, and the resulting `magnetic' string tension turns out to be
\beq
\sigma_{k,N}=\frac{\Lambda^2}{\lambda}\,\frac{N}{\pi}\,\sin\frac{\pi k}{N}\,,
\la{sigmaM-k}\eeq
which coincides with the `electric' string tension \ur{sigma-k} found
from the correlators of the Polyakov lines, for all $k$-strings!

Several comments are in order here.
\begin{itemize}
\item The `electric' and `magnetic' string tensions should coincide only in the
limit $T\to 0$ where the Euclidean $O(4)$ symmetry is restored. Both calculations
have been in fact performed in that limit as we have ignored the temperature-dependent
perturbative potential \ur{Ppert}. If it is included, the `electric' and `magnetic'
string tensions split.
\item despite that the theory \ur{Z4} is 3-dimensional, with the temperature entering
just as a parameter in the Lagrangian, it ``knows'' about the restoration of
Euclidean $O(4)$ symmetry at $T\to 0$.
\item the `electric' and `magnetic' string tensions are technically obtained in
very different ways: the first is related to the mass of the elementary excitation
of the dual fields ${\rm w}_m$, whereas the latter is related to the mass of the dual field
soliton.
\end{itemize}

\section{Cancelation of gluons in the confinement phase}

To prove confinement, it is insufficient to demonstrate the area law for large
Wilson loops and the zero average for the Polyakov line: it must be shown that
there are no massless gluons left in the spectrum. We give an argument
that this indeed happens in the dyon vacuum.

A manifestation of massless gluons in perturbation theory is the Stefan--Boltzmann
law for the free energy:
\beq
-\frac{T}{V}\log {\cal Z}_{\rm SB} =\frac{F_{\rm SB}}{V}
=-\frac{\pi^2}{45}\,T^4\,(N^2-1).
\la{FSB}\eeq
It is proportional to the number of gluons $N^2\!-\!1$ and has the $T^4$ behavior
characteristic of massless particles. In the confinement phase, neither is permissible:
If only glueballs are left in the spectrum the free energy must be ${\cal O}(N^0)$
and the temperature dependence must be very weak until $T\approx T_c$ where it abruptly
rises owing to the excitation of many glueballs.

The nonperturbative free energy corresponding to the minimum of the dyon-induced
potential energy as function of the holonomy \ur{calP2} is
\beq
\frac{F_{\rm dyon}}{V}=-\frac{N^2}{2\pi^2}\,\frac{\Lambda^4}{\lambda^2}.
\la{Fdyon}\eeq
It is ${\cal O}(N^2)$ but temperature-independent. We have doubled the minimum from \eq{calP2}
keeping in mind that there are also anti-dyons and assuming that their interactions
with dyons is not as strong as the interactions between dyons and anti-dyons
separately, as induced by the determinant measure \ur{G}, therefore treating
dyons and anti-dyons as two independent ``liquids''. (By the same logic, the string
tension \ur{sigma-k} has to be multiplied by $\sqrt{2}$ as due to anti-dyons.)

Dyons force the system to have the ``most nontrivial'' holonomy \ur{muconf}.
For that holonomy, the perturbative potential energy \ur{Ppert} is at its maximum equal to
\beq
\frac{F_{\rm pert,\, max}}{V}=\frac{\pi^2}{45}\,T^4\,\left(N^2-\frac{1}{N^2}\right).
\la{Fmax}\eeq
The full free energy is the sum of the three terms above.

We see that the leading ${\cal O}(N^2)$ term in the Stefan--Boltzmann law is {\em canceled}
by the potential energy precisely at the confining holonomy point and nowhere else!
In fact it seems to be the only way how ${\cal O}(N^2)$ massless gluons can be
canceled out of the free energy, and the main question shifts to why does the
system prefer the ``most nontrivial'' holonomy. Dyons seem to answer that question.

\section{Deconfinement phase transition}

\begin{center}
\begin{table}[h]
\hspace{0.4cm}
\begin{tabular}{|c|c|c|c|c|c|}
\hline
&$SU(3)$ & $SU(4)$ & $SU(6)$ & $SU(8)$ & $N\to\infty $ \\
\hline
&&&&&\\
$T_c/\sqrt{\sigma}$, {\rm theory}& 0.6430 & 0.6150 & 0.5967 & 0.5906 
& $0.5830+\frac{0.4795}{N^2}+\frac{0.5006}{N^4}+...$\\
&&&&&\\
\hline
&&&&&\\
$T_c/\sqrt{\sigma}$, {\rm lattice}& 0.6462(30) & 0.6344(81) & 0.6101(51) 
& 0.5928(107) & $0.5970(38)+\frac{0.449(29)}{N^2}\;({\rm fit})$\\
&&&&&\\
\hline
\end{tabular}
\end{table}
\end{center}

As the temperature rises, the perturbative free energy grows as $T^4$ and eventually it
overcomes the negative nonperturbative free energy \ur{Fdyon}. At this point, the trivial
holonomy for which both the perturbative and nonperturbative free energy are zero,
becomes favorable. Therefore an estimate of the critical deconfinement temperature
comes from equating the sum of \Eq{Fdyon} and \Eq{Fmax} to zero, which gives
\beq
T_c^4=\frac{45}{2\pi^4}\,\frac{N^4}{N^4-1}\,
\frac{\Lambda^4}{\lambda^2}\,.
\la{Tc}\eeq
As expected, it is stable in $N$. A more robust quantity, both from the theoretical
and lattice viewpoints, is the ratio $T_c/\sqrt{\sigma}$ where $\sigma$ is the
string tension in the fundamental representation, since in this ratio
the poorly known parameters $\Lambda$ and $\lambda$ cancel out:
\beq
\frac{T_c}{\sqrt{\sigma}}=\left(\frac{45}{4\pi^4}\,\frac{\pi^2 N^2}{(N^4-1)\sin^2\frac{\pi}{N}}
\right)^{\frac{1}{4}}\quad\stackrel{N\to\infty}{\longrightarrow}\quad
\frac{1}{\pi}\left(\frac{45}{4}\right)^{\frac{1}{4}}.
\la{ratio}\eeq

In the Table, we compare the values from \Eq{ratio} to those measured in lattice simulations
of the pure $SU(N)$ gauge theories~\cite{Teper}; there is a surprisingly good agreement.
A detailed study of the thermodynamics of the phase transition will be published elsewhere.

\section{An exceptional gift}

It is illuminating to go beyond the $SU(N)$ gauge groups and consider the YM theory based
on {\it e.g.} the exceptional group $G(2)$. It has rank 2 and is similar to the $SU(3)$ group
but has only a trivial center, meaning that only a unity element commutes with all other
group elements (in $SU(N)$ there are $N$ such matrices being $N$-roots of unity; they form
the group center $Z_N$). Since confinement in $SU(N)$ is often associated with the
symmetric distribution of the Polyakov line eigenvalues with respect to the $Z_N$ permutations,
and the deconfinement transition is associated with the spontaneous breaking of $Z_N$ symmetry,
while neither is a feature of $G(2)$, questions have been raised whether there is at all confinement
in $G(2)$, leave alone the confinement-deconfinement phase transition.

However, recent numerical simulations of $G(2)$ performed by three groups~\cite{G2}
showed that there is confinement at low $T$ and a first-order deconfinement transition. By
confinement we mean here the zero average of the Polyakov line in the lowest 7-dimensional
representation. These findings pose difficulties for the confinement scenarios based on
the center of the group, {\it e.g.} on center vortices.

The dyon scenario works, however, exceptionally well it this case, too. We have repeated all the
steps described in this paper but applied to the $G(2)$ group. There are three types
of fundamental dyons, like in $SU(3)$, but neutral clusters are formed by four;
overall neutrality requires that one type of dyons have to enter twice more often than the
two others. The minimum of the free energy for the dyon ensemble lies exactly at the holonomy
corresponding to the zero Polyakov line. At a critical temperature there is a first-order 
transition to a phase with a nonzero average Polyakov line~\cite{DP-G2}.

We stress that these results are sensitive to the dynamics as they are not in the least
enforced by symmetry. They provide strong support to the dyon scenario of confinement.
It should be added that instantons with nontrivial holonomy and dyons have been directly
observed in lattice simulations~\cite{HU-ITEP}.

\section{Summary}

What happens in the semiclassical approximation based on dyons, can be summarized as follows:
\begin{itemize}
\item The ensemble of dyons favors dynamically the confining
value of the holonomy. This is almost clear, given that the weight is proportional
to the product of individual actions of fundamental dyons, and it is maximal when
the actions are equal. Such holonomy corresponds to the zero of the Polyakov line
\item Dyons form a sort of Coulomb plasma (but an exactly solvable variant of it)
with an appearance of the Debye mass both for ``electric'' and ``magnetic'' (dual) photons.
The first gives rise to the exponential fall-off of the correlation of two Polyakov lines,
{\it i.e.} to the linear heavy-quark potential, the second yields the area law for spatial
Wilson loops
\item ${\cal O}(N^2)$ massless gluons cancel out from the free energy, and only massive (string?)
excitations are left.
\end{itemize}

The reason why a semiclassical approximation works well for strong interactions
(where all dimensionless quantities are, generally speaking, of the order of unity) is
not altogether clear. A possible justification has been outlined in the Introduction:
After UV renormalization is performed about the classical saddle points and the scale parameter
$\Lambda$ appears as the result of the dimensional transmutation, further quantum corrections
to the saddle point are a series in the running 't Hooft coupling $\lambda$ whose argument
is typically the largest scale in the theory, in this case ${\rm max}(T, n^{1/4})$ where
$n$ is the $4d$ density of dyons. An estimate shows that the running $\lambda$ is
between $1/4$ at zero temperature and $1/7$ or less at critical temperature. Therefore,
although these numbers are ``of the order of unity'', in practical terms they indicate
that high order loop corrections are not too large. Let us recall that quite an accurate
computation of anomalous dimensions in critical phenomena from the $\epsilon$-expansion
by Fisher and Wilson~\cite{FW} is based on truncating the Taylor expansion in $\epsilon$
at the first couple of terms, where $\epsilon\!=\!1$ or sometimes 2 !\footnote{One of us
(D.D.) takes the opportunity to thank Michael Fisher and Valery Pokrovsky for a discussion
of this numerical miracle.}

Unfortunately, approximations made in Ref.~\cite{DP-07} and reproduced above are not
limited to neglecting higher loop corrections. We have (i) ignored dyon interactions induced
by the small oscillation determinant over nonzero modes (although we did
take into account that it renormalizes the gauge coupling giving rise to the
scale parameter $\Lambda$, and that it leads to the perturbative potential energy
as function of the holonomy), (ii) neglected the interactions of dyons of opposite
duality, treating them as two noninteracting ``liquids'', (iii) conjectured a simple
form of the dyon measure which may be incorrect when two {\em same-kind} dyons
come close. Although certain justification for these approximations can be put
forward~\cite{DP-07} it is desirable not to use them at all, and that may be possible.

\begin{theacknowledgments}
This work has been supported in part by Russian Government grants RFBR-06-02-16786
and RSGSS-5788.2006.2.
\end{theacknowledgments}

\end{document}